# A systematic evaluation of Silicon-rich Nitride Electro-optic Modulator design and tradeoffs

ALEX FRIEDMAN,[1,*] DMITRII BELOGOLOVSKII,[1] ANDREW GRIECO, [1] AND YESHAIAHU FAINMAN [1]

[1]*Peer Review, Publications Department, Optica, 2010 Massachusetts Avenue NW, Washington, DC 20036, USA*

**amfriedm@eng.ucsd.edu*

**Abstract:** We present a study of linearized $\chi^{(3)}$ based electro-optic modulation beginning with an analysis of the nonlinear polarizability, and how to linearize a modulator based on the quadratic third order DC-Kerr effect. Then we perform a numerical study, designing a linearized $\chi^{(3)}$ phase modulator utilizing Silicon-rich Nitride where we show that a phase modulator with a $V_\pi L_\pi$ metric of 1 Vcm or a $V_\pi L_\pi \alpha$ metric of 37VdB is achievable and a $V_\pi L_\pi$ as low as 0.5Vcm in a push-pull Mach Zehnder Interferometer. This numerical study argues that linearized modulation exploiting the $\chi^{(3)}$, and $\chi^{(2)}$ as applicable, is possible and can allow for high-speed modulation using a CMOS compatible material platform.

## 1. Introduction

Optical interconnects form a major part of the disruptive impact of integrated optical systems in a variety of applications, and therefore have driven continued interest in finding the next generation of optical modulators. Historically, high-speed optical modulators have relied upon lithium niobate [1,2] where thanks to the lithium niobate on insulator platform $V_\pi L_\pi$ metrics on the order of 1.8Vcm have been achieved [3,4], while in search of higher efficiencies other high-k dielectrics have been considered such as barium titanate [5]. All of these materials have three primary challenges:(i) they are in general not CMOS compatible making fabrication more expensive compared to a CMOS process which can be done by a foundry, (ii) they have low refractive indices compared to Silicon requiring larger waveguide cross-sections to achieve reasonable mode-confinement, and (iii) they all have higher RF permittivity leading to smaller electric fields in the same cladding at the same applied voltage. As a result many optical interconnects still utilize carrier dispersion in silicon waveguides ( $V_\pi L_\pi \cong 3.5\ Vcm,$ with an insertion loss of 9.2dB ) [6], where excess propagation loss due to high dopant concentrations can limit performance, and so there remains interest in a CMOS compatible alternative to such techniques that can disrupt optical modulators in CMOS manufacturing. The issue remains, however, that most CMOS compatible materials either do not exhibit a $\chi^{(2)}$, or exhibit a negligible small one such as lower index silicon nitride films. In recent years work in literature, including past work by the author's [7], has demonstrated that not only can Silicon-rich Nitride films exhibit a non-zero $\chi^{(2)}$ but that their refractive index [8, 9], thermo-optic coefficient [10], as well as $\chi^{(2)}$ and $\chi^{(3)}$ [11] are all enhanced with increasing silicon content and that this is true even in the case of low temperature plasma-enhanced chemical vapor deposition-based SRN films [8,12]. In this manuscript, we will undertake a systematic evaluation of the contributions from second and third order nonlinearities in arbitrary materials and make a case for a $\chi^{(3)}$ based linearized electro-optic modulator utilizing a form of heterodyne gain. We will then conduct a numerical study of such a modulator, designing phase-type modulator based on our past works Silicon-rich Nitride film properties, achieving $V_\pi L_\pi$ metrics from 2 to 3.5Vcm and $V_\pi L_\pi \alpha$ metrics of 116VdB as a phase modulator, or as low as 1Vcm in a push-pull Mach Zehnder Interferometer intensity modulator. We will then explore the integration of such a phase-shift element into a ring resonator cavity as an intensity modulator demonstrating that with proper cavity design and allowing for a degree of coupling miss-match extinction ratios between 12dB and 20dB are achievable. Finally, we will conclude with some discussion on the inherent tradeoffs with such a design and argue that a linearized $\chi^{(3)}$ based modulator, can serve as a viable CMOS compatible alternative to use materials lacking $\chi^{(2)}$ nonlinearities, and as a pure phase modulator alternative to traditional plasma-dispersion approaches in silicon.

## 2. Nonlinear Polarizability Analysis

We begin with a brief analysis of second and third order nonlinear optical effects in nonlinear materials with emphasis on the presence of an applied external electric field. Here we include consideration for effects of higher order nonlinearities, beyond that of $\chi^{(3)}$ alone, as research has shown that most CMOS compatible materials lack a $\chi^{(2)}$ as a result of their crystal symmetry [13-16]. Although this is an abbreviated discussion on the topic, a useful place to start is through the induced polarization which can be written as follows [17]:

$$P(r,t) = \epsilon_0 \left[\chi^{(1)} E(r,t) + \chi^{(2)} E^2(r,t) + \chi^{(3)} E^3(r,t) + \dots\right] \quad (1)$$

In equation 1, $\chi^{(1)}$, $\chi^{(2)}$ and $\chi^{(3)}$ represent the first, second and third order nonlinear susceptibilities respectively and are treated as tensors of rank two, three, and four, respectively. This is a useful formalism because both modulation and wave mixing are understood as solutions to the nonlinear form of Maxwell's equation [17]. If we allow the total electric field $E(r,t)$to be a sum of an optical wave ($E_\omega$) and applied electric field (electro-static $E_{dc}$ and time-varying $E_{ac}$), we can derive expressions for the contributions to the nonlinear portion of the induced polarization, grouping and simplifying terms based on their contributions. Below in equation 2, as an example, we show three first terms of eq. 1 expansion, grouped and labeled with various nonlinear effects.

$$\begin{aligned}\bar{P}_{NL} &= \epsilon_0 \{ \overbrace{\chi^{(2)}[2EE^* + E_{dc}^2]}^{\text{Electrostatic}} + \overbrace{\chi^{(2)}[2E_{dc}Ee^{-j\omega t} + 2E_{dc}E^* e^{j\omega t}]}^{\text{Pockels}} + \overbrace{\chi^{(2)}[E^2 e^{-j2\omega t} + E^{*2} e^{j2\omega t}]}^{\text{SHG}} + \overbrace{\chi^{(3)}[E_{dc}^3 + 6EE^*E_{dc}]}^{\text{Electrostatic}} \\ &+ \underbrace{\chi^{(3)}[(3E_{dc}^2 E + 3E^2E^*)e^{-j\omega t} + (3E_{dc}^2 E^* + 3EE^{*2})e^{j\omega t}]}_{\text{Kerr effect and DC-Induced Pockels}} + \underbrace{\chi^{(3)}[3E_{dc}E^2 e^{-j2\omega t} + 3E_{dc}E^{*2} e^{j2\omega t}]}_{\text{EFISHG}} + \underbrace{\chi^{(3)}[E^3 e^{-j3\omega t} + E^{*3} e^{j3\omega t}]}_{\text{THG}} \}\end{aligned} \quad (2)$$

This represents a subset of the possible terms in the nonlinear polarizability, due to our specific choice of terms in the total electric field and additionally the tensorial notation has been suppressed here for simplicity. The terms present in this expansion relate to the various forms of nonlinear processes that utilize the susceptibility as shown in table 1. Table 1 shows a breakdown of the terms of the nonlinear polarization based on the combination of optical wave and applied field present in the induced polarization. The items in this table can be thought of as falling into one of two different categories: (1) Effects at the fundamental optical frequency – Switching and modulation and (2) Effects at harmonic frequencies (including 0 frequency) – wave mixing. Switching and Modulation based on the nonlinear susceptibility is typically thought to comprise of the Pockels effect and the Quadratic Electro-optic effect, sometimes called the DC-Kerr effect, where-as wave mixing comprises three and four wave mixing, active and passive. However, terms such as the "EFI-SHG" (here EFI stands for electric field induced) and "Modulated SHG" blur this distinction and can allow methods for both enhancing second order processes as well as analyzing third order nonlinearities via four-wave mixing.

**Table 1: A breakdown of the induced polarization**

■ $\chi^{(1)}$
■ $\chi^{(2)}$
■ $\chi^{(3)}$
■ $\chi^{(4)}$

| | | Applied Field | | | |
|---|---|---|---|---|---|
| | | **0** | **$E^{dc}$** | **$E^{ac}$** | **$E^{ac}E^{dc}$** |
| **Optical Wave** | **0** | | $\chi^{(2)}$ Optical rectification | $\chi^{(3)}$ Optical rectification | |
| | **$\omega$** | Fundamental Wave | Pockels Effect | | DC-Kerr Effect |
| | **$2\omega$** | SHG | EFI-SHG | Modulated SHG | DC-$\chi^{(4)}$ Effect |
| | **$3\omega$** | THG | EFI-THG | Modulated THG | ⋱ |

Importantly, if we allow for a combination of electro-static $E_{dc}$ and time-varying $E_{ac}$ together with the optical $E_{\omega}$, we can derive a term (last column in table 1) which allows for third order nonlinearity based linear modulation. This is especially interesting to explore, as unlike $\chi^{(2)}$ which is dependent on crystal structure, $\chi^{(3)}$ is present in all materials. In the remainder of this manuscript we will only consider non-resonant electronic nonlinearities as this type of nonlinearity can respond at ultra-fast speeds and is thus of interest for high-speed modulation, as well as be useful for wave-mixing applications [18]. In the following we consider the case of an arbitrary material which has some set of $\chi^{(2)}$ and $\chi^{(3)}$ tensors, under the presence of an applied electric field which has both an electro-static (Edc) and a time-varying ac term (Eac). From this we derive an expression for each combination to the change in refractive index showing them below in table 2 based on $\chi^{(2)}$ vs $\chi^{(3)}$ and their combination of Edc and Eac fields.

**Table 2: Second and Third order contributions to the change in refractive index**

| | DC | AC | AC+DC |
|---|---|---|---|
| $\boldsymbol{\chi^{(2)}}$ | $\Delta n_{dc}^{\chi^{(2)}} = \sum_{jk} \frac{\chi_{ijk}^{(2)}}{n_k^{eq}} E_j^{dc}$ | $\Delta n_{ac}^{\chi^{(2)}} = \sum_{jk} \frac{\chi_{ijk}^{(2)}}{n_k^{eq}} E_j^{ac}$ | Not Applicable |
| $\boldsymbol{\chi^{(3)}}$ | $\Delta n_{dc}^{\chi^{(3)}} = \sum_{jk} \frac{3\chi_{ijjk}^{(3)}}{2n_k^{eq}} E_j^{dc^2}$ | $\Delta n_{ac}^{\chi^{(3)}} = \sum_{jk} \frac{3\chi_{ijjk}^{(3)}}{2n_k^{eq}} E_j^{ac^2}$ | $\Delta n_{ac+dc}^{\chi^{(3)}} = \sum_{jk} \frac{3\chi_{ijjk}^{(3)}}{n_k^{eq}} E_j^{ac} E_j^{dc}$ |

Table 2 reveals a few interesting features of such an arbitrary material. Specifically, for such a material under the presence of an Edc and Eac field there will be of course a static "bias" change in refractive index represented by $\Delta n_{dc} = \Delta n_{dc}^{\chi^{(2)}} + \Delta n_{dc}^{\chi^{(3)}}$; however, there will also be a modulated component of the change in refractive index represented by $\Delta n_{ac} = \Delta n_{ac}^{\chi^{(2)}} + \Delta n_{ac}^{\chi^{(3)}} + \Delta n_{ac+dc}^{\chi^{(3)}}$. If we are interested in constructing a linearized modulator utilizing a given materials $\chi^{(3)}$, as well as $\chi^{(2)}$ if it has it, then the $\Delta n_{ac}$ term is the important one to analyze. Additionally one can notice that the $\Delta n_{ac+dc}^{\chi^{(3)}}$ mixed term has an extra factor of 2 compared with the non-mixed terms, $\Delta n_{dc}^{\chi^{(3)}}$ and $\Delta n_{ac}^{\chi^{(3)}}$. From these formulas and Table 2 the only contributing term that is not linear in Eac is the $\Delta n_{ac}^{\chi^{(3)}}$ term, which is the reason third order modulation is typically quadratically chirped, a problem for modulator design; however, the term $\Delta n_{ac+dc}^{\chi^{(3)}}$ has two benefits. First, if we require the Edc >> Eac, a condition that will dictate the degree of linearity, then the $\Delta n_{ac+dc}^{\chi^{(3)}}$ term is approximately linear in Eac, and additionally under such a condition it is naturally true that $\Delta n_{ac}^{\chi^{(3)}} \ll \Delta n_{ac+dc}^{\chi^{(3)}}$ which allows us to ignore the quadratically chirped term.

Secondly, the $\Delta n_{ac+dc}^{\chi^{(3)}}$ term exhibits a natural form of what can be thought of as a heterodyne gain. This term has a weak high-frequency field (Eac) and a strong low-frequency field (Edc) the product of which produce an effect at the high-frequency field's enhanced by the strength of the low-frequency field (Edc). While this will require the Edc field strength to be high, it will allow the Eac field strength to be proportionally lower, this can be a solution of interest in the CMOS case as 10's of volts-dc can be acceptable whereas the AC voltage is the one that needs to be as low as possible, even sub 1V in some cases. Figure 1 shows an example of how by controlling the ratio of Eac to Edc the quadratic chirping in the resulting change in phase can by removed. Note that as the Eac/Edc ratio increases so does the peak phase change because the magnitude of Eac increases.

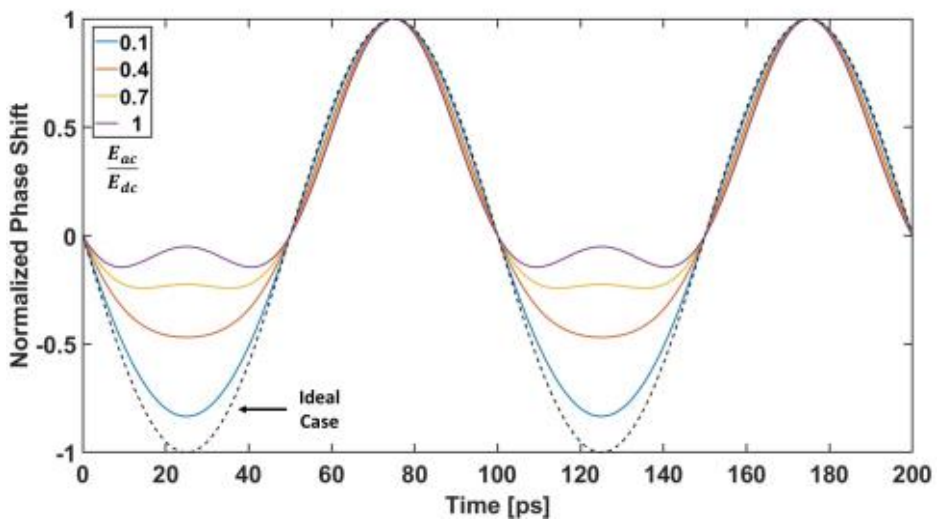

Fig. 1. An example for a 10Ghz modulated Eac wave as a function of the fraction of Eac/Edc for a fixed Edc of 1.22x10$^8$ V/m. In the above plot the black dashed line represents the idealized case.

It is important to discuss the trade-off between $\chi^{(2)}$ based Pockels modulation and this $\chi^{(3)}$ based DC-Induced Kerr modulation. As has been derived previously in the literature there is a general rule that the order of magnitude of $\chi^{(2)}$ and $\chi^{(3)}$ should be expected to be $\frac{\chi^{(1)}}{E_{at}}$ and $\frac{\chi^{(1)}}{E_{at}^2}$ respectively [17], where $E_{at}$ is the atomic electric field strength. This has the important implication that we should expect effects based on $\chi^{(3)}$ to be weaker than that of $\chi^{(2)}$ because the order of magnitude of their coefficients in general differ by the atomic electric field strength and thus an effective $\chi^{(2)}$ induced by the presence of an applied dc electric field could only approach that of the expected inherent $\chi^{(2)}$ when the applied electric field approaches that of $E_{at}$, in materials whose $\chi^{(2)}$ originates from a crystal lattice dipole. Such a condition is of course not possible as the breakdown field of a given material will in general be much lower than $E_{at}$ meaning that we will reach the maximum strength of applied electric field before the combination of $\chi^{(3)}E_{applied}$ is expected to be of the order of the inherent $\chi^{(2)}$. While this may indeed be a limitation, in realistic materials, especially CMOS compatible materials, the $\chi^{(2)}$ tensor is often zero due to crystal symmetry, in such cases this technique can still be useful as all materials have a non-zero $\chi^{(3)}$ tensor. In silicon-rich nitride specifically the story becomes more interesting as the exact origin of the non-zero $\chi^{(2)}$ in silicon-rich nitride is still not well understood. As a result this general rule may not hold true for such a material. For example, if we define $\chi_{eff}^{(2)} = 3\chi^{(3)}E_{dc}$ and consider our previous results in [7] then we find that at Edc = 1x10$^8$ V/m we have a $\chi_{eff}^{(2)}$ of 180 pm/V compared to the measured 14 pm/V inherent $\chi^{(2)}$ which indicates that a value of $\chi_{eff}^{(2)} > \chi^{(2)}$ is possible. For this reason, we will explore the design of a linearized modulator based on the DC-Induced Kerr effect in silicon-rich nitride. It has been shown in literature in past work by the author [7] as well as others [11] that PECVD silicon-rich nitride can exhibit refractive indices as high as 3.1 and has a comparably enhanced $\chi^{(3)}$. Additionally, silicon-rich nitride films are expected to have a high breakdown field, as silicon nitride films can exhibit very high breakdown fields, as well as low optical loss over a broader spectral range that Silicon [9, 11, 19]. As the example calculation of a $\chi^{(2)}_{eff}$ shows, these features can make silicon-rich nitride a very attractive candidate.

## 3. Phase Modulator Design

We begin by analyzing the phase shifter element on its own, as a fundamental building block as well as a phase-type modulator of its own. Figure 2(a) shows a schematic cross-section of the proposed device structure. In the structure a Silicon-rich Nitride waveguide is located on a $SiO_2$ buried oxide layer. We then create a conformal thin dielectric shield layer (e.g., deposited using ALD) followed by construction of gold (Au) electrodes. Finally, the structure has a top cladding layer of $SiO_2$. This structure can be thought of as the generic case of a device which lacks such a shield layer, the case where the shield dielectric is $SiO_2$ and a case where the left and right electrodes could also be placed directly onto the bottom $SiO_2$ layer and then the center electrode formed after depositing a desired thickness of $SiO_2$. We consider the generic design case as presented as a useful layout for discussing important tradeoffs of such designs. Firstly, the objective is to maximize the strength of the applied electric field within the waveguide core while minimizing the induced optical loss, a trade-off which will dictate the optimal device performance regime. One key parameter that will dictate the strength of the applied field is the ratio of the relative permittivity of the shield layer to that of the Silicon-rich Nitride waveguide layer.

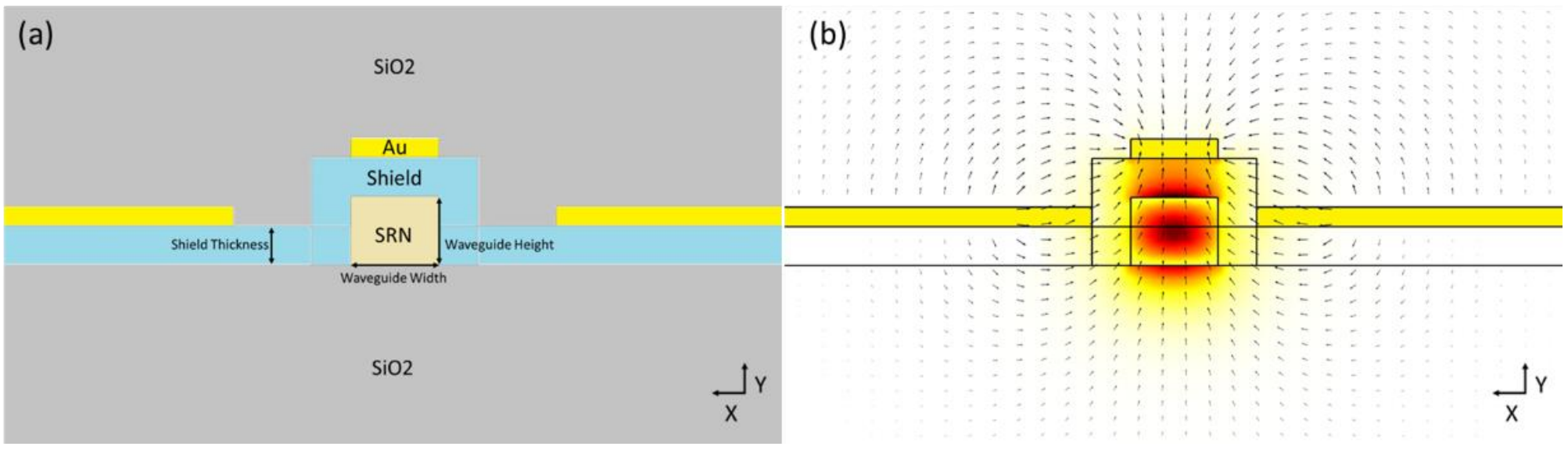

Fig. 2. (a) Schematic cross-section of the proposed device structure. A Silicon-rich Nitride waveguide sits on a SiO2 buried oxide layer. On top of the Silicon-rich Nitride waveguide is a thin dielectric shield layer onto which Ground-Signal-Ground gold electrodes are formed. Finally, the top of the structure is top clad with $SiO_2$. (b) Image showing the TM optical mode for a 350nm thick, 450nm wide SRN waveguide along with field lines for an applied electric field from the GSG field lines.

Therefore, if we utilize a thin cladding layer of silicon nitride, which is known to have an RF permittivity around 7.2 [20], to more closely match the RF permittivity of the interface to that of the SRN waveguide core which we have previously measured to be 9.44 [7], we can increase the penetration of the applied electric field into the waveguide. However, the trade-off is that utilizing a silicon nitride shield layer will reduce the refractive index contrast of the waveguide core, reducing mode confinement and thus increasing induced optical loss.

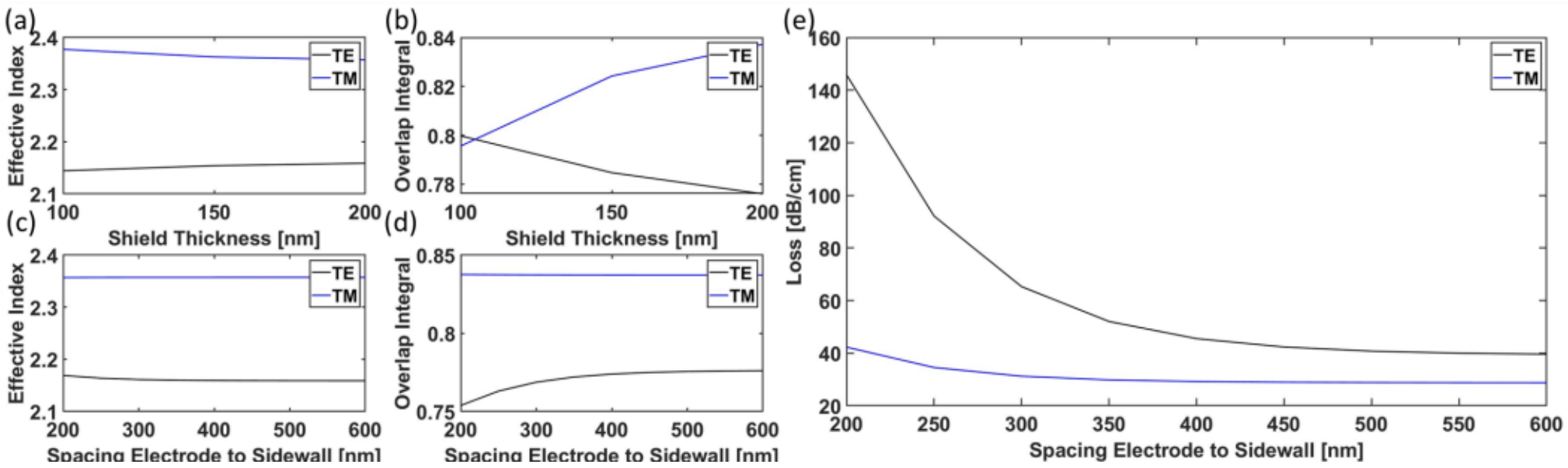

Fig. 3. (a) Effective Index of the TE-like and TM-like modes vs SiO2 ‘shield’ thickness. (b) Overlap Integral of the TE-like and TM-like modes with the SRN waveguide core versus SiO2 ‘shield’

thickness. (c) Effective Index of the TE-like and TM-like modes vs spacing from the electrode to waveguide sidewall. (d) Overlap Integral of the TE-like and TM-like modes with the SRN waveguide core versus spacing from the electrode to waveguide sidewall. (e) Propagation Loss vs electrode to sidewall spacing for the TE-like and TM-like modes.

In general the usage of an intermediate dielectric shield layer between the waveguide core and the metal electrodes is a necessity due to optical losses from metals; however, once introduced the mismatch in RF permittivity between the intermediate dielectric shield layer and the waveguide core will "shield" the higher RF permittivity waveguide core from the applied electric fields reducing the strength of the field within the nonlinear medium. The solution then is to utilize a material which matches the RF permittivity of the silicon-rich nitride core; however, when considering practical materials often the RF permittivity and the refractive index increase in tandem. For example, at ~7.2 silicon-nitride [20] has an expected RF permittivity closer to that of silicon-rich nitride layer however it has a higher refractive index at 1.8 to 1.95 compared to the 1.45 of $SiO_2$ which has an RF permittivity in the range of 3.75 to 4.45 [21, 22]. The result of this is that in a realistic CMOS compatible material stack with limited choices for dielectric shield layers there is a tradeoff between the strength of the applied electric field and the loss of the optical mode from the lower modal confinement of a higher refractive index cladding layer. In the design of the phase shift element in this study we will use $SiO_2$ as the shield layer in order to mitigate excess loss from modal deconfinement. This means that since the shield layer is the same material as the cladding it serves as a physical spacer rather than any additional RF permittivity matching. Our proposed structure will be based on the expected $\chi^{(2)}$ and $\chi^{(3)}$ values for silicon-rich nitride films of similar refractive index in literature such as 14 pm/V and $6x10^{-19}$ $m^2/v^2$ from our past work [7] with a good review in [23]; we will utilize the measured RF permittivity from that work as well [7]. Figure 2(b) shows a COMSOL model of the TM optical mode for a 350nm thick 450nm wide Silicon-rich Nitride waveguide along with the applied electric field lines from the ground-signal-ground (GSG) electrodes. Utilizing COMSOL and Lumerical we simulate the electrical and optical fields of the structure as a function of both the electrode to waveguide sidewall spacing as well as the thickness of the shield layer and optimize the width and height of the waveguide to minimize excess at 500nm thick and 350nm wide and assume a $SiO_2$ shield layer. As in this case the shield layer is assumed to be the same as the cladding, it serves simply as a physical spacer rather than as both a physical spacer and permittivity matching. For these simulations as outlined in section 2, we will consider the up to the case where the mean value of $E_{dc}$ in the core is as large as $1.22x10^8$ V/m at peak value, taking results from COMSOL simulations.

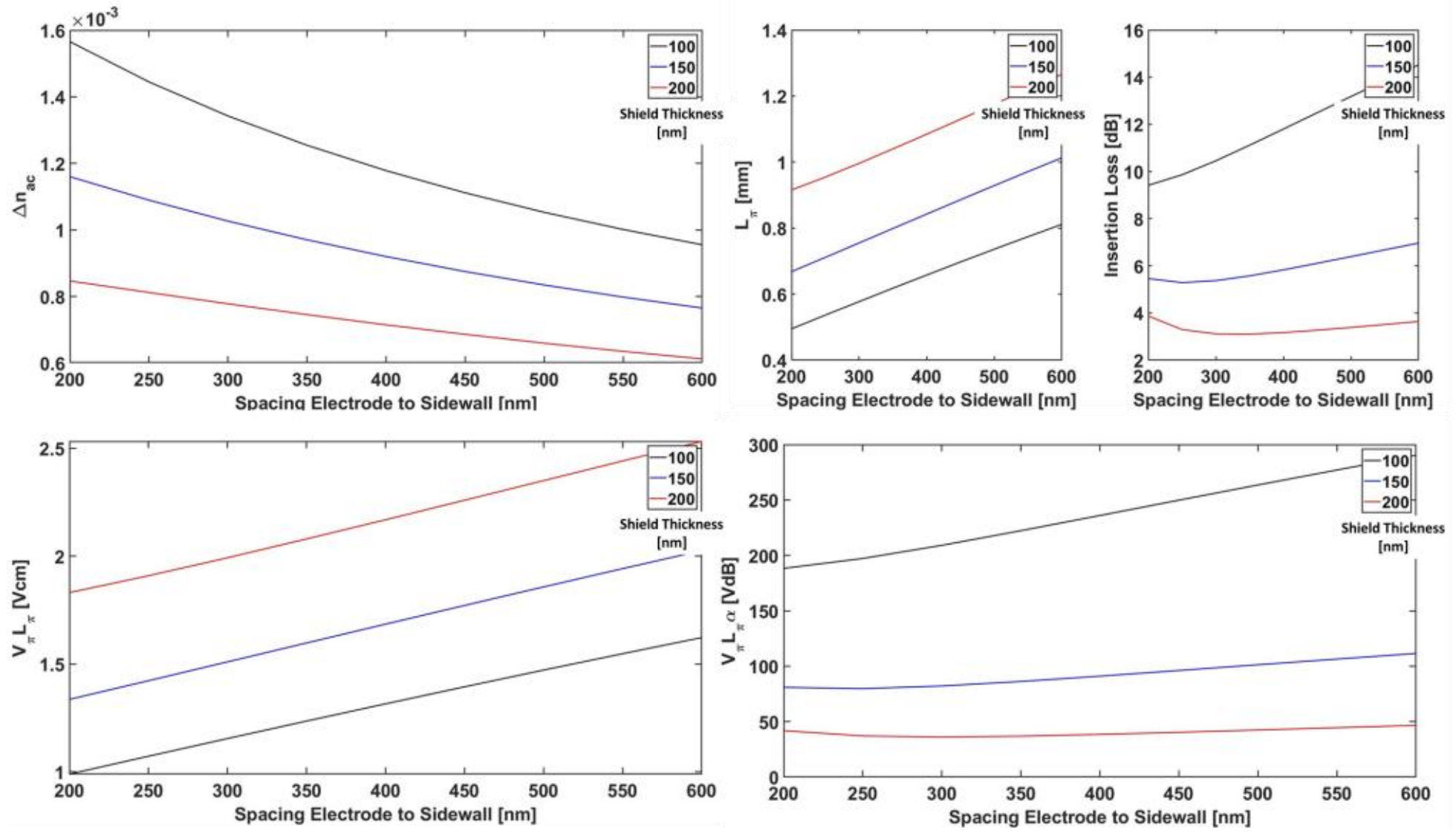

Fig. 4. (a) A plot of the change in refractive index under the presence of a combination of AC and DC field. In such a situation if we require that Edc >> Eac then we can write this as approximately $\Delta n_{ac} = \Delta n_{ac}^{\chi^{(2)}} + \Delta n_{ac+dc}^{\chi^{(3)}}$. (b) A plot of the require length for a $\pi$ phase shift based on $\Delta n_{ac}$ as a function of electrode to sidewall spacing and shield thickness. (c) A plot of lumped loss for a phase shifter with a length of $L_\pi$ as a function of electrode to sidewall spacing and shield thickness. (d) From the change in refractive index we can determine the metric $V_\pi L_\pi$. (e) While the metric $V_\pi L_\pi$ is a commonly used metric for such devices, a more comprehensive metric is $V_\pi L_\pi \alpha$ which includes loss and is thus in units of V-dB instead of V-cm.

Figure 3(a-b) shows the effective index and overlap integral of the TE-like and TM-like modes as a function of shield thickness. In the case where the shield is simply the same as the cladding it serves as a physical spacer predominately for the central electrode. As a result, at small shield thickness the TM-like optical mode gets pulled into the thin gap between the center electrode and the waveguide core, increasing the effective index but decreasing the overlap with the SRN layer. In figure 3(c-d) a similar effect occurs for the TE-like mode when the electrodes are brought closer to the sidewall of the waveguide with the enhancement of the effective index, and corresponding decrease in overlap integral, being smaller due to large gaps than the shield layer thickness for the TM-like mode. The tradeoff of course is loss, which is shown in Figure 3(e), whereas the electrode to sidewall spacing is reduced the loss increases. Naturally, the TE-like mode sees a faster increase in loss from bringing the electrodes closer to the side-wall were as the TM-like case will see a faster increase as the shield thickness decreases. It is important to remember the tensorial nature of $\chi^{(2)}$ and $\chi^{(3)}$ in light of these facts, specifically the different tensor components utilized depending on the orientation of the optical polarization and applied electric field. As discussed in section 2, the fundamental relation between second and third order nonlinearities, and the presences of a non-zero $\chi^{(2)}$ in SRN, means that all else being equal a TM polarized optical mode and vertical applied field will produce the largest change in refractive index as it will utilize both the largest $\chi^{(2)}$ and $\chi^{(3)}$ tensor. As such in this manuscript we will focus on the case utilizing the $\chi^{(3)}_{3333}$ tensor component which is the case for a TM-like optical mode and a vertical applied field; however, a variety of other configurations could be explored in the future such as combinations which use not-colinear tensor components for example an in-plane applied field and a TM-like optical mode. By modeling of the applied electric field we predicted the $\Delta n_{ac}$, shown in figure 4(a), for electrode to sidewall spacings from 200nm to 600nm and shield thickness from 100nm to 200nm. The trade-off here is clear, at a fixed voltage smaller electrode to sidewall spacings result in larger applied field strengths at a fixed voltage, and thus larger changes in refractive index. Similarly, decreasing the shield layer thickness increases the applied field strength in the waveguide core at a fixed voltage and thus increases the change in refractive index. From the change in refractive index curves the minimum length required for a $\pi$ phase shift ($L_\pi$) is determined as a function of electrode to sidewall spacing and shield thickness in figure 4(b). As the shield thickness increases the strength of the applied electric field in the waveguide core reduces thus requiring longer path lengths to maintain a $\pi$ phase shift. Using the lengths from figure 4(b) and the propagation loss values discussed in figure 3(e) we generated predicted values for the insertion loss in figure 4(c) reaching values as low as 3.1dB. Figure 4(d) then, shows the predicted $V_\pi L_\pi$ for each corresponding case under the condition that $\frac{E_{dc}}{E_{ac}} \cong 10$. From this it is clear that we can achieve competitive $V_\pi L_\pi$ metrics, from 1 to 1.8Vcm. However, this is only a part of the story as a comprehensive figure of merit should include the loss, so we define an additional figure of merit to be $V_\pi L_\pi \alpha$ which results in a unit of VdB. Figure 4(e) shows such a figure of merit including the loss in the analysis. The way to interpret such a figure of merit is to consider that at a $V_\pi L_\pi \alpha$ of 37 VdB a 20Vpp AC voltage would have an insertion loss of 1.85dB. It is through a combination of these two figures of merit that the design space, and tradeoffs between voltage, length, and insertion loss can be understood. Based on these results then the following performance can be achieved for such a Silicon-rich Nitride Modulator.

**Table 3: Example Possible Design parameters**

| Vdc [V] | Vac [Vpp] | L [mm] | IL [dB] |
|---|---|---|---|
| -200 | -20 | 1 | 3.1 |
| -200 | -6.2 | 3.2 | 10 |

Table 3 shows the expected performance for our considered design. A clear trend in this design then is a tradeoff between voltage and lumped loss, for example at a $V_\pi L_\pi$ of 2Vcm with a 6.2Vpp ac voltage we get 10dB or at 20Vpp we get 3.1dB. On the other hand we can achieve the minimum 1Vcm $V_\pi L_\pi$ metric at an insertion loss of 9.4dB.

## 4. Intensity Modulator Design

In the previous section we discussed the design and performance of a phase modulator, which provides important insight into the fundamental performance of the underlying device. We will now discuss how such a phase shift element can be used as an intensity modulator by embedding it into a ring resonator cavity or in a Mach Zehnder interferometer (MZI) configuration. Both ring-resonator and Mach Zehnder configurations have their own advantages and drawbacks which we will discuss but they can be thought of as broadly representing the two categories of intensity modulators, resonant and non-resonant respectively. As has been well reported in literature [24, 25] a ring resonator can be viewed as a form of a resonant filter, which when the resonant condition, $\phi_{roundtrip} = m2\pi$, is satisfied light is lost from the transmission port while off resonant light is allowed to pass. It is this condition that allows a phase-modulator embedded into the cavity of a ring resonator to be realized as an intensity modulator, the phase introduced by the phase modulator adds to the nominal roundtrip phase and changes the wavelength at which the nominal round trip phase plus the phase change from the modulator results in an integer multiple of $2\pi$. Equation 3 below shows the well-known expression for the transmission from an all-pass configuration ring-resonator where $r$ is the self-coupling coefficient of the bus, $k$ is the cross-coupling coefficient, $a$ is the single-pass amplitude transmission, and p is the single-pass phase shift [24]:

$$T_p = \frac{a^2 - 2racos(\phi) + r^2}{1 - 2arcos(\phi) + (ra)^2} \quad (3)$$

Therefore, in the idealized case the critical coupling condition can be shown to be when the coupled power is equal to the power lost in the ring, or $r = a$. In figure 5 below we consider our phase-shift element from section 3 embedded into the cavity of a silicon-rich nitride 45$\mu$m bend radius ring resonator. Additionally, we consider the case where there is an 'unintended' mismatch between the amplitude transmission coefficient and the self-coupling coefficient of 0.05, as a typical value to account for fabrication and design variation from past experience and where the phase modulator comprises 90% of the cavity length, to accommodate the coupling region.

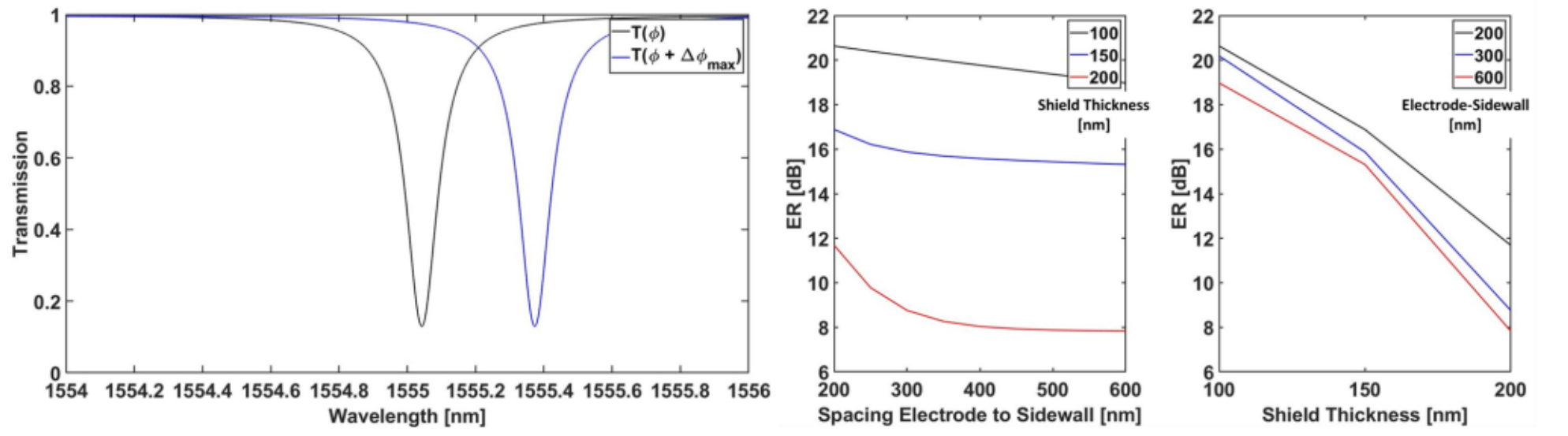

Fig. 5. (a) Simulated Transmission spectra and maximum shift in transmission spectra for a 45um bend radius ring modulator assuming a $\Delta_{r-a} = 0.05$ mismatch between the single pass amplitude transmission coefficient and the self-coupling coefficient of the bus. (b) Extinction Ratio as a function of electrode to sidewall spacing for shield thickness 100nm to 200nm. (c) Extinction Ratio as a function of shield thickness for electrode-sidewall spacings from 200nm to 600nm.

Figure 5(a) shows the transmission spectra of the nominal device $T(\phi)$ in black along the expected shifted transmission spectra $T(\phi + \Delta\phi_{max})$ due to $\Delta n_{ac}$. Figure 5(b) and(c) then shows the expected extinction ratio (ER), here defined as the ratio transmission from the nominal spectra and the transmission under the peak phase change, as a function of electrode to sidewall spacing and shield thickness. Two general trends can be seen in figure's 5(b) and (c), increasing electrode to sidewall spacing decreases the ER and second increasing shield thickness decreases the ER. As long as the ring resonator is in the critical coupling regime, increasing either the electrode to sidewall spacing or the shield thickness reduces the ER by decreasing the strength of the applied field within the waveguide core. On the other hand, as was discussed in section 3, reducing the electrode to sidewall spacing or the shield thickness increases the loss, in the case of a ring resonator this impacts the maximum achievable quality factor. In turn a commonly used definition of the quality factor is the ratio of stored energy in the cavity to energy dissipated per cycle and therefore is a measurement of the rate of decay of energy in the cavity [26]. These two factors form a tradeoff which is visible in Figure 6(a) and (b).

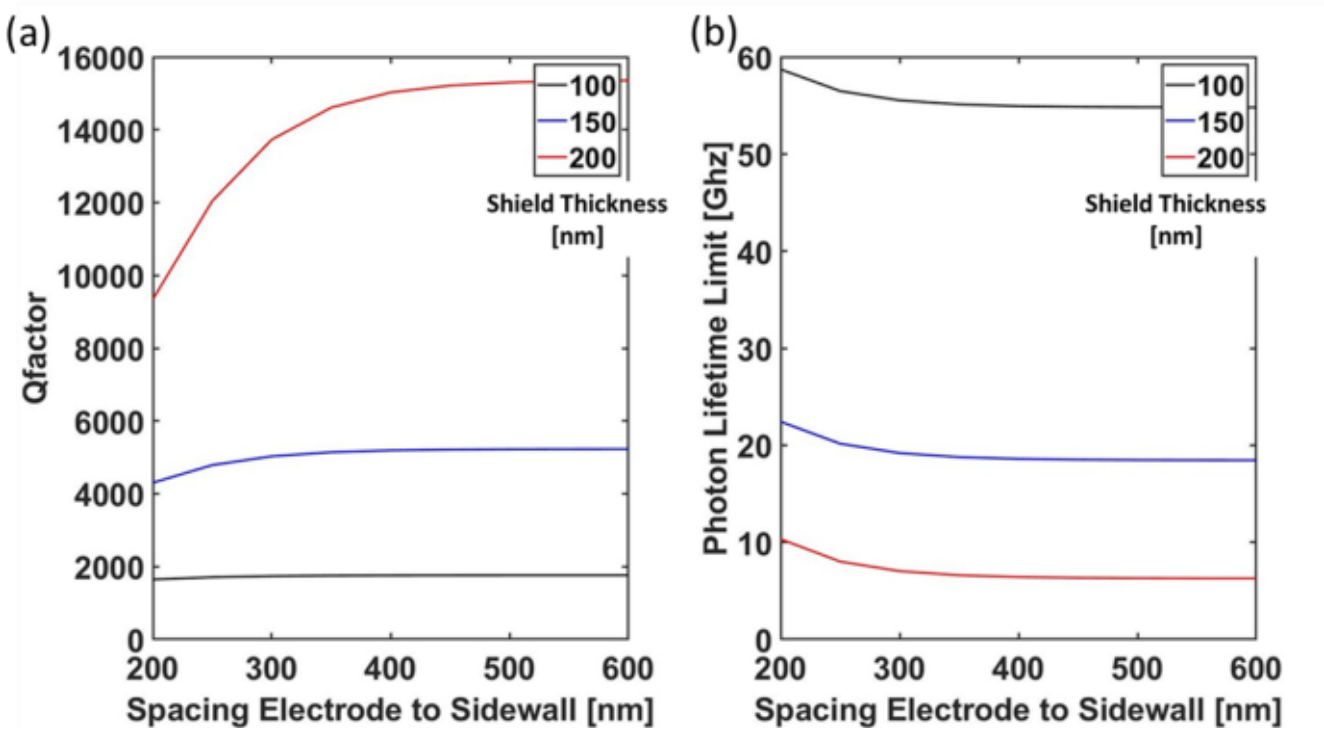

Fig. 6. (a) Quality Factor versus electrode to sidewall spacing for shield thickness 100nm to 200nm (b) Photon lifetime limited bandwidth vs electrode to sidewall spacing for shield thickness 100nm to 200nm.

Figure 6(a) above shows the expected quality factor of a ring resonator with a $\Delta_{r-a}$ mismatch of 0.05 vs electrode to sidewall spacings for shield thickness from 100nm to 200nm. As mentioned above, the reduction in quality factor for increasing loss (decreased electrode to sidewall spacing or decreased shield thickness) can clearly be seen. As ring resonators are resonant cavities the enhanced intensity contrast is a tradeoff with the ring-down time of the cavity which limits response times. Using the quality factor vs electrode to sidewall spacing and shield thickness we present the photon-lifetime limited bandwidth, $\frac{1}{2\pi\tau_{cavity}}$, where $\tau_{cavity}$ is the photon lifetime of the cavity and related to the quality factor as $Q = \frac{\omega_0 \tau_{cavity}}{2}$ [26, 27]. From figure 6(b) then we can see that increasing quality factor's results in lower photon lifetime limited bandwidths, reaching as low as ~10Ghz for the largest quality factors. Additionally, there is a slight enhancement in photon lifetime limited bandwidth at the smallest electrode to side wall spacings. Based on these results a photon lifetime limit of 60Ghz requires a quality factor of 2000, which is naturally achieved at a shield thickness of 100nm. Therefore, integrating our phase modulator into a ring resonator cavity will allow ER's of 10dB to 20dB and photon lifetime bandwidths of 60Ghz for Q factors of 2000. The non-resonant alternative we will discuss here is the Mach Zehnder Interferometer configuration. Unlike the ring resonator, the MZI configuration being non-resonant does not have a photon lifetime limit to its bandwidth instead being limited by the capacitive load of the electrodes. Here we consider an unbalanced MZI, with a mismatch length of 200$\mu$m, and the phase modulator of length $L_\pi$ from section 3 in both arms. Figure 7(a) shows the simulated spectral response of such an unbalanced MZI in the nominal case $T(\phi)$.

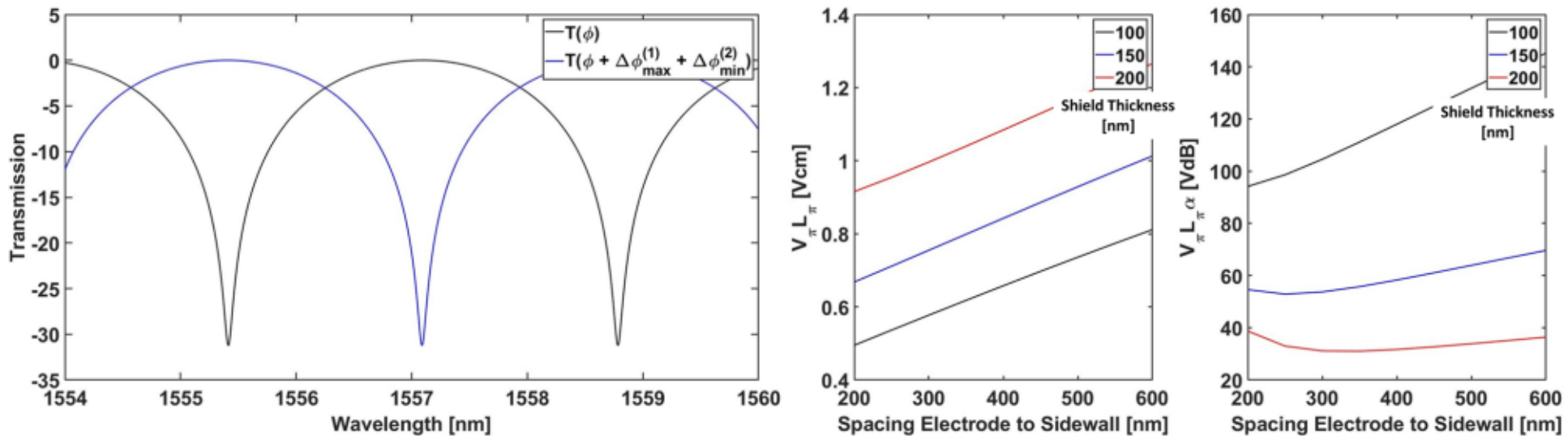

Fig. 7. (a) Simulated Transmission spectra and maximum shift in transmission spectra for a MZI with an imbalance length of 200$\mu$m along with electrodes of the $L_\pi$ length from section 3 in both arms driven in push-pull. (b) A plot of the $V_\pi L_\pi$ metric versus electrode to sidewall spacing and shield thickness. (c) A plot of the $V_\pi L_\pi \alpha$ metric versus electrode to sidewall spacing and shield thickness.

In the case of the MZI by controlling the relative phases introduced into both arms we can form an intensity modulator. In this case we will consider the push-pull configuration where the electrical driving signal of the two arms is $\pi$ out of phase with each other. The resulting $V_\pi L_\pi$ and $V_\pi L_\pi \alpha$ metrics can be seen in figures 7(b) and (c) respectively. The analysis here is a fairly straightforward extension of the phase modulator, as figure 7(a) shows a full $\pi$ phase shift in the spectra is achievable and the corresponding figures of merit clarify that the in addition to converting the phase modulator into an intensity modulator, we have cut the $V_\pi L_\pi$ in half, achieving values of 0.5Vcm to 0.9Vcm as a result of utilizing push pull. The primary trade-off in this design is one of compactness, unlike the ring resonator configuration which can be made as small as twice the bend radius of choice, the MZI here requires electrode lengths of $L_\pi$ which are 1mm, or longer, in this case. The trade-off here then is you cut the $V_\pi L_\pi$ metric down as low as 0.5Vcm in the push-pull MZI; however, the longer electrode's relative to the ring resonator as well as being driven in push-pull, which is likely to lead to a parallel set of capacitances, relative to the phase modulator means that it will have the largest capacitance of the three configurations. If we compare such results to literature, such as that of lithium niobate on insulator devices [3], we find that SRN Mach Zehnder can achieve a $V_\pi L_\pi$ between 0.5 to 0.9Vcm while lithium niobate on insulator devices achieves values in the range 1.8Vcm; however, if we compare to only CMOS compatible techniques such as a depletion type silicon modulator [6] we find values in the range of 3.5Vcm. In table 4 below we have summarized a comparison of various modulator platforms from literature.

Table 4: A comparison of Modulator Performance

| | $V_\pi L_\pi$ [Vcm] | IL [dB] | RF Permittivity | Bandwidth | CMOS | Reference |
|---|---|---|---|---|---|---|
| *This work* | 1 | 1.55 | 9.45 | 20 to 60Ghz | *yes* | - |
| Depletion type Silicon | 3.5 | 9.2 | N/A | ~20Ghz | yes | [6] |
| Lithium Niobate on Insulator | 1.8 | 2 | 28 | ~ 15 Ghz | no | [3] |
| Hybrid Lithium Niobate on Silicon | 2.7 | 1.8 | not stated | > 70 Ghz | no | [4] |
| Silicon on BTO thin-film | 0.23 | not stated | 100 - 3000 | ~20 Ghz | no | [5] |

These techniques range from lithium niobate on insulator to hybrid Silicon on barium titanate (BTO) thin-film approaches. Of these various approaches the silicon on BTO thin-film achieves the clear best $V_\pi L_\pi$ metric; however, being a silicon on BTO thin-film device requires post-processing and is not in general CMOS compatible. Additionally, the large nonlinearities that allow for low voltage's are in general smaller when used for wave-mixing. Of the approaches in the table, our result and the depletion type silicon modulator are the only two that can be clearly defined as CMOS compatible material stacks. Our numerical study shows that SRN DC-Induced Kerr modulators can achieve competitive $V_\pi L_\pi$ metrics and being a low temperature PECVD process it can bring new capabilities to CMOS compatible platforms.

## 5. Discussion and Conclusion

Traditionally electro-optic modulators have relied upon second order nonlinearities, utilizing the Pockels effect; however, materials that exhibit non-zero $\chi^{(2)}$ tensors are generally not CMOS compatible. Meanwhile $\chi^{(3)}$ based modulation has typically been seen as un-attractive due to a much weaker nonlinearity exhibited by most materials as well as the quadratic nature of the effect. In this manuscript we have undertaken a systematic evaluation of electro-optic nonlinearities in a generic material and then made the case that the unique combination of $\chi^{(2)}$ and $\chi^{(3)}$ exhibited by SRN makes it a good candidate for capacitive $\chi^{(3)}$ based electro-optic modulation. We have shown that SRN can achieve $V_\pi L_\pi$ metrics as low as 1Vcm in a MZI configuration and extinction ratios as high as 18dB in a ring resonator configuration all utilizing a cmos compatible material platform. Additionally, we addressed the traditional drawback of quadratic chirping in $\chi^{(3)}$ based modulators by showing that proper choice of the Eac/Ec ratio can not only linearize the change in phase but can also be seen as a heterodyne gain approach with the mixing of the weak high frequency Eac term and the strong low frequency Edc term. While for some applications utilizing a non-CMOS device such as a lithium niobate on insulator modulator can be acceptable, there is a need for CMOS compatible alternatives to such devices. As it stands now if a designer is limited to CMOS processing due to a desire to utilize cost effective tapeouts then they are primarily limited to carrier dispersion-based modulators in silicon. In this manuscript we have argued that adoption of a $\chi^{(3)}$ based modulator can provide additional utility to such CMOS platforms and that silicon-rich nitride is a good candidate for such adoption. PECVD based silicon nitride films are already widely utilized in CMOS tapeouts, and as has been shown in the past by the authors [7, 8], with proper tuning of gas flow ratios a high refractive index PECVD silicon-rich nitride film can be achieved under otherwise the same processing conditions. In this manuscript we have shown that the unique advantages of high confinement guiding in a low RF permittivity high $\chi^{(3)}$ and low loss in such a platform makes it an attractive candidate for integration into standard CMOS process flows. Further exploration of linearized $\chi^{(3)}$ based modulators in a variety of material platforms can provide new and unique capabilities and deserves further investigation.

**Funding.** This work was supported by the Defense Advanced Research Projects Agency (DARPA) DSO NLM and NAC programs, the Office of Naval Research (ONR), the National Science Foundation (NSF) grants DMR-1707641, CBET-1704085, NSF ECCS-180789, NSF ECCS-190184, NSF ECCS-2023730, NSF ECCS-190184, the Army Research Office (ARO), the San Diego Nanotechnology Infrastructure (SDNI) supported by the NSF National Nanotechnology Coordinated Infrastructure (grant ECCS-2025752ECCS-1542148), the Quantum Materials for Energy Efficient Neuromorphic Computing-an Energy Frontier Research Center funded by the U.S. Department of Energy (DOE) Office of Science, Basic Energy Sciences under award # DE-SC0019273; Advanced Research Projects Agency Energy (LEED: A Lightwave Energy-Efficient Datacenter), and the Cymer Corporation.

**Acknowledgments.** We thank all of UCSD's nano3 cleanroom staff and Dr Maribel Montero for their assistance with sample fabrication.

**Disclosures.** The authors declare no conflicts of interest.